# Experimental investigation of surface instability of a thin layer of a magnetic fluid

Arthur Zakinyan, Levon Mkrtchyan, Yuri Dikansky

Department of Physics, Institute of Mathematics and Natural Sciences, North Caucasus Federal University, 1 Pushkin Street, 355009 Stavropol, Russia

Address correspondence to A. Zakinyan:
phone: +7-8652-57-00-33,
postal address is presented above,
e-mail: zakinyan.a.r@mail.ru

**Abstract.** In the present work the instability of a flat horizontal thin layer of a magnetic fluid (the depth of no more than 50 µm) under the action of a uniform magnetic field is studied experimentally. It was revealed that the development of instability under the action of tilted magnetic field can lead to the formation of parallel ridges on a fluid surface; the ridges undergo a transformation into hexagonal system of conical peaks with the magnetic field increasing. The necessary conditions for the formation of these surface patterns are studied. It was found that the development of instability of quite thin layers may result in layers breakup. The dependencies of instability wave number on the system physical parameters are obtained. The time for the development of instability is measured. The experimental results are compared with the existing theory and discussed.

## 1. Introduction

The matters of instability of fluid interfaces and formation of surface patterns under the action of external fields are widely studied and still receive a great deal of attention in the fluid mechanics range of problems. Apart from independent interest, the results of such studies can be useful in the understanding of a number of natural phenomena and in technical applications. Among the problems of fluid instabilities the problem of instability of a free surface of magnetic fluid in a dc magnetic field is widely known. A magnetic fluid is a colloidal suspension of ultra-fine ferro- or ferri-magnetic nanoparticles suspended in a carrier fluid. When the external dc vertical magnetic field reaches some critical value, the horizontal magnetic fluid surface spontaneously deforms into regularly spaced conical peaks, usually distributed on a hexagonal lattice. It is so-called normal field or Rosensweig instability; see [1] for a review. The available theoretical and experimental publications devoted to the Rosensweig instability mainly deal with the case of a deep magnetic fluid layer. At the same time the investigations of the instability of layers of the thickness comparable or less than the characteristic linear scale (capillary length) are also important. Periodic structures of pointed drops formed as a result of decomposition of thin layers are of interest in the designing of devices for injecting charged particles with the help of magnetic field [2, 3]. The instability of the finite thickness layers in a perpendicular magnetic field has been theoretically considered in some classical works [4, 5]. Peculiarities of thin layers instability is also a subject of several recent theoretical researches [6–8]. As to experiments in this area, it seems that only works [9–12] can be mentioned. The dependencies of the critical magnetic field strength and critical wave number on the thickness of the magnetic fluid layer have been measured. The supercritical wave number dependencies on the magnetic field strength and layer thickness have been obtained [9–12]. It was shown that the thickness dependence of the instability phenomena is the most pronounced in a layer of the thickness below 100 μm. However, the characteristic time for the development of instability has not been analyzed in the existing experiments.

While a wide variety of detailed investigations of the instability of a magnetic fluid surface in a perpendicular magnetic field are available, the data on the behavior of a magnetic fluid surface in a tilted magnetic field are scarce. The presence of a tangential component of a magnetic field changes the character of instability and pattern of a magnetic fluid surface qualitatively. The problem of the magnetic fluid instability in a tilted magnetic field has been studied theoretically in [13–18]. It has been shown that the tangential component of a magnetic field stabilizes a certain range of harmonic perturbations propagating along it, and it leads to the emergence of the surface pattern in the form of parallel stripe-like ridges with axes parallel to the tangential component of a magnetic field. It should be noted that several assumptions with loss of generality have been made in the mentioned theoretical studies.

A first experimental observation of liquid ridges on a magnetic fluid surface in a tilted magnetic field was presented in [19]. However, no further experiments have been reported. More recently, rather more detailed experimental studies of the problem have been attempted [20, 21]. It was demonstrated that, when increasing the tilted field strength, the flat surface gives way to liquid ridges, and a further increase results in a transition to a pattern of stretched hexagons. The threshold values of the magnetic field at which these patterns are formed were measured as a function of a tilt angle. The magnetic fluid layers investigated in the mentioned works are quite thick, and it can be considered as if they are of quasi-infinite depth. The dependence of the surface patterns peculiarities on the layer thickness and the wave number of instability in a tilted magnetic field were not examined in the existing experimental works.

In the present work we study the instability of a thin layer of a magnetic fluid in an arbitrary orientated external uniform dc magnetic field. Our experiments were performed with 5- to 50-μm-thick layers.

## 2. Theoretical background

The problem of the instability of a thin layer of a magnetic fluid in an arbitrary orientated magnetic field has been studied theoretically by Korovin [15]. We will use the results of [15] to analyze our experimental data and to compare the measured results with the theoretical predictions. A layer of a motionless magnetic fluid with a flat free surface on a horizontal nonmagnetic plate immersed in a uniform tilted magnetic field has been studied in [15]. The layer has been considered to be thin in the sense that the condition $h/\lambda \ll 1$ is satisfied. Here $h$ is the layer thickness and $\lambda$ is the wave length. The case of $\tau_d/\tau_i \ll 1$ was analyzed. Here $\tau_i$ is the characteristic time for the development of instability, $\tau_d = h^2/\nu$ is the characteristic time for the diffusion of vorticity across the fluid layer, and $\nu$ is the kinematic viscosity of the fluid. Our experimental conditions are in general agreement with the model requirements.

Within the scope of a linear perturbation analysis the following dispersion equation has been obtained:

$$s = \frac{h^3}{3\eta}\left\{-\rho g\left(k_1^2 + k_2^2\right) + \frac{\mu_0}{2}\left[M_{03}^2\left(k_1^2 + k_2^2\right)^{3/2} \right.\right.$$
$$\left.\left. - M_{01}^2 k_1^2 \left(k_1^2 + k_2^2\right)^{1/2}\right] - \sigma\left(k_1^2 + k_2^2\right)^2\right\} \quad (1)$$

where $s$ is the imaginary part of the perturbation frequency $\omega$, which is purely imaginary quantity; $\eta$ is the dynamic viscosity of the fluid; $g$ is the acceleration of gravity; $\rho$ is the fluid density; $\sigma$ is the fluid surface tension; $\mu_0$ is the magnetic constant; $M_{03}$ is the vertical component of the magnetization of magnetic fluid; $M_{01}$ is the projection of the magnetization of magnetic fluid onto the direction of tangential component of external magnetic field; $k_1$ and $k_2$ are the wave numbers of perturbations propagating along the tangential component of magnetic field and perpendicularly to it. The magnetization of magnetic fluid is determined from the experimental magnetization curve $M(H_i)$ ($H_i$ is the magnetic field inside the magnetic fluid). Field strength inside the fluid, $H_i$, is computed from the externally applied magnetic field $H_e$ by the solution of an implicit equations:

$$H_{i3} = H_{e3} - NM_{03}(H_{i3}), \quad H_{i1} = H_{e1},$$

where $N$ is the demagnetizing factor, which was set equal to unity for the thin flat layer under study. The analysis of equation (1) shows that the tangential component of a magnetic field stabilizes the harmonic perturbations propagating along it.

The system of equations

$$\partial s/\partial k_2 = 0, \quad s = 0 \quad (2)$$

is the condition of the instability onset. Having numerically analyzed equations (2) with regard to the magnetization curve taken experimentally, one can determine the critical magnetic field strength and critical wave number.

It should be noted that the rate of a magnetic field rise is an important factor for the surface pattern formation process. When the slowly rising magnetic field reaches the critical value, the surface pattern with critical wave number appears. If the magnetic field increases further the wave number of the pattern will remain nearly the same, only wave amplitude must grows. But the pattern wave number may change if the rate of a magnetic field rise is greater than the instability increment, i.e., if the magnetic field settles down before the surface instability starts developing. Thus, if the magnetic field of the strength greater than critical value is switched on very rapidly the wave number of surface pattern is no longer a constant and depends on the magnetic field strength [9]. For the supercritical magnetic fields only the first equation of the system (2) should be considered to find the wave number of the most unstable perturbations (supercritical wave number).

In the case of the supercritical magnetic field from the dispersion equation (1) one can find the characteristic time for the development of the most rapidly growing harmonic (the instability development time) [15]:

$$\tau_i = 2\pi \cdot \left[\max s(k_1, k_2)\right]^{-1}. \quad (3)$$

The presented equations (1)–(3) will be used to compare our experimental data with the results of calculation. The figures below show, where possible, the both experimental and theoretical results.

## 3. Materials and methods

A layer of a motionless magnetic fluid with a flat free surface on a horizontal hard nonmagnetic plate placed in a dc uniform tilted magnetic field is studied. In our experiments we used two kerosene-based magnetic fluids with dispersed magnetite nanoparticles of about 10 nm diameter stabilized with oleic acid. The properties of the first magnetic fluid (MF-1) are: density is 1700 kg/m$^3$, surface tension coefficient at the fluid–air interface is 0.028 N/m, dynamic viscosity is 20 mPa·s, and saturation magnetization is 76 kA/m. The properties of the second magnetic fluid (MF-2) are: density is 1400 kg/m$^3$, surface tension is 0.027 N/m, dynamic viscosity is 15 mPa·s, and saturation magnetization is 55 kA/m. The magnetization curves of magnetic fluids have been obtained by the use of a vibrating-coil magnetometer (Fig. 1).

We used a following experimental procedure. A layer of the magnetic fluid applied on a rectangular glass plate (with surface area $S = 5$ cm$^2$) was placed between the electromagnet poles in a dc uniform magnetic field. Thickness, $h$, of the layer was determined from known volume, $V$, of the fluid distributed over the surface of the plate ($h = V/S$). To ensure the thickness homogeneity the layer was subjected to mechanical vibrations for 10 s. The electromagnet could be rotated through 90° to set an arbitrary angle between the magnetic field direction and the normal to the layer surface (a tilt angle, $\theta$). The magnetic field was increased gradually and the visual examinations of the surface patterns resulting from fluid–field interaction were performed.

Increasing the external magnetic field strength, $H_e$, we observed a transition from the flat layer of magnetic fluid to the pattern of liquid ridges. The wave vector of the pattern was oriented perpendicularly to the horizontal field component. The value of the external magnetic field strength at which these ridges become discernible was measured (the first critical field, $H_{c1}$). The further increase in field strength leads to the growth of the ridges amplitude and can result in the layer breakup into the regularly spaced parallel liquid stripes if the layer thickness is sufficiently small. The typical pattern of liquid ridges is presented in Fig. 2. The observed patterns are nearly uniform over the surface of the layer; this means that edge effects can be ignored under our experimental conditions (since the wave length of the instability is small compared with the sizes of the layer plane). Increasing the magnetic field strength further leads to a breakup of liquid ridges into conical peaks arranged into a hexagonal pattern. This hexagonal pattern of peaks is analogous to the pattern arising in a normal magnetic field [9–12], but in this case the conical peaks are inclined to layer plane and the hexagons are stretched along the horizontal field component. The value of the external magnetic field strength at which the transition to a hexagonal pattern of peaks takes place was measured (the second critical field, $H_{c2}$).

Next we have investigated the wave number of the pattern of ridges. It was difficult to measure the critical wave number, $k_c$, in our experiments, and we have measured the supercritical wave number, $k$, of the pattern which appears when a magnetic fluid layer is subjected to a supercritical magnetic field (the field strength is more than $H_{c1}$ and less than $H_{c2}$). In our experiments the magnetic field settling time was about 0.06 s that is less than the instability build-up time for sufficiently thin layers. Therefore, the supercritical wave number can be reasonably measured. To measure the pattern wave number the magnetic fluid layer subjected to a magnetic field was dried and then examined under an optical microscope. During drying (~15 min), the pattern wave number did not change. By measuring the distance between the centers of neighboring ridges, the pattern wave length, $\lambda$, was determined and wave number, $k = 2\pi/\lambda$, was obtained.

In the case of the supercritical magnetic fields the time for the development of instability has been measured. The measurements have been performed at the tilt angle $\theta = 0$, i.e., at the magnetic field oriented perpendicularly to the layer. The instability development process has been recorded by means of a high-speed digital camera at 300 frames per second. The moment of the magnetic field onset has been indicated by the light flash from the light-emitting diode connected to the electromagnet coils and actuated by the electric current generating the magnetic field. It was observed that after the magnetic field switching on the layer surface remains flat for some period of time and then transforms into the pattern of peaks. The duration of this time period has been measured as the instability development time.

## 4. Experimental results

We have investigated the angular and thickness dependences of the critical magnetic field strength. The MF-1 has been used in these experiments. Fig. 3 shows the measured data for the transition to ridges, $H_{c1}$, and the transition to hexagons, $H_{c2}$. These curves increase monotonically. The transition from flat surface to ridges does not demonstrate the hysteretic behavior in our experiments. On the contrary, the transition to hexagons has a slightly hysteretic character. In the present study we will discuss the transition to hexagons upon increase of magnetic field. As seen from Fig. 3, at small tilt angles, $\theta$, both dependences coalesce, and $H_{c1}$ cannot be distinguished from $H_{c2}$. In this case only the hexagonal pattern of peaks arises omitting the pattern of ridges. Fig. 3 demonstrates the ranges of an external magnetic field strength and a tilt angle in which the different surface configurations appear: flat surface, parallel ridges and conical peaks. Each of the experimental dependences has been measured three times, and the obtained results are different from each other by no more than 5%. Fig. 3 also shows the theoretical dependence of the critical external magnetic field strength, $H_{c1}$, on the tilt angle. The comparison with the corresponding results of [20] for the thick layer shows that the instability of thin layer begins at higher field strength despite the fact that we have used the magnetic fluid with stronger magnetic properties than that used in [20]. It can be concluded that thin layer is more stable than thick layer.

It should be noted that no transition to a hexagonal pattern of peaks is implied by the theoretical analysis of [15]. But the bifurcation to hexagons has been predicted using a perturbative energy minimization procedure in [17]. It was shown in [17] that for broken symmetry ridges always precede hexagons. This theory considers the infinite layer of inviscid magnetic fluid with linear magnetization law and restricted to magnetic permeabilities $\mu < 1.4$.

It has been found that the critical values of the magnetic field $H_{c1}$ and $H_{c2}$ at which the magnetic fluid layer becomes unstable depend on its thickness. Fig. 4 shows the experimental dependences of $H_{c1}$ and $H_{c2}$ on the thickness of the layer. These curves decrease monotonically. It confirms the conclusion that thin layers are more stable. As it can be seen from Fig. 4, for the case of sufficiently thick layers the values of $H_{c1}$ and $H_{c2}$ are not much different from each other, and it becomes difficult to distinguish the pattern of ridges as the instability leads to the formation of pattern of peaks. The dependence of the critical field on the layer thickness is inconsistent with the theoretical analysis of [15] according to which the critical magnetic field is thickness independent. This is due to the assumption of small layer thickness ($h/\lambda << 1$) made in the theoretical analyses. The tilted field instability of magnetic fluid layer of arbitrary depth has been theoretically studied in [13]. The analysis of dispersion equation obtained in [13] shows the decrease of critical magnetic field with the layer thickness increasing, which is in a qualitative agreement with our experiments. But the theoretical results of [13] demonstrate a very large quantitative disagreement with our experimental data. This is due to the neglect of fluid viscosity and assumption of linear magnetization law made in [13]. For this reason, we do not use the theoretical results of [13] in the present work.

We have obtained the dependences of the pattern supercritical wave number on the magnetic field strength, on the layer thickness, and on the tilt angle. The MF-1 has been used in these experiments. In Fig. 5 the dependence of the pattern wave number on the external magnetic field strength, $H_e$, is presented. It is seen that the wave number rises monotonically with the field strength. Fig. 6 shows the dependence of the pattern wave number on the tilt angle. As can be seen, the wave number and the tilt angle are inversely related. Figs. 5 and 6 also demonstrate the theoretical dependences of the instability wave number on the external magnetic field strength and on the tilt angle extracted from Eqs. (1) and (2). In Fig. 7, the wave number of the pattern is plotted against the thickness of the magnetic fluid layer. It is seen that the wave number decreases with the thickness increasing. The corresponding calculation according to [15] gives the thickness independence of the supercritical wave number and contradicts our experimental observations. The thickness dependence of the supercritical wave number has been theoretically demonstrated for the normal field instability [4, 5, 9], but the existing theoretical studies of the tilted field instability does not predict such dependence.

It should be noted that the wave number of the tilted field instability has not been analyzed in the existing experimental works performed with thick layers [19–21]. Furthermore, the instability wave length in [19–21] was of the order of the horizontal sizes of vessel with magnetic fluid, this leads to the essential boundary effect on the experimental results. On the contrary, in our experiments with thin layers the wave length was by two orders of magnitude smaller than the layer horizontal size.

The time for the development of instability has been measured. The MF-2 has been used in these experiments. Fig. 8 shows the experimental dependence of the instability development time, $\tau_i$, on the thickness of the layer. As can be seen, the instability development time decreases monotonically with the thickness increasing. Fig. 9 shows the $\tau_i$ dependence on the external magnetic field strength. It is seen that the instability development time decreases with the field strength. The corresponding theoretical results are also presented in Figs. 8 and 9.

As it can be seen, the trend of the theoretical and experimental curves presented in Figs. 3, 5, 6, 8, and 9 is qualitatively the same. On the other hand, the comparison of experimental and theoretical data demonstrates some quantitative disagreement. In addition to the above-mentioned simplified assumption made in the theoretical analysis, this disagreement may be associated with nonlinear effects and with adhesion of the magnetic fluid to the substrate.

## 5. Conclusions

Thus, our experimental investigation revealed new aspects in the development of instability of a thin layer of a magnetic fluid placed in a tilted magnetic field, namely: the thickness dependences of the threshold values of a magnetic field; the instability supercritical wave number dependences on the field strength, the tilt angle, and the layer thickness; the instability development time dependencies on the field strength and the layer thickness. A comparison of some experimental and theoretical results shows that they are in a qualitative agreement. At the same time, the obtained results for the thickness dependences of the critical magnetic field and the supercritical wave number contradict the theoretical predictions. Unfortunately, there is a lack of full theoretical studies of the tilted field instability of viscous magnetic fluid layer of arbitrary depth with nonlinear magnetization law. It should also be noted that the effects of breakup of a thin layer and transition to a pattern of peaks with magnetic field increasing were not fully theoretically analyzed in the existing works and call for further studies in this field.

It can be expected that the results obtained here will be useful in the investigation of pattern formation processes in such phenomena as non-Boussinesq inclined layer convection, magnetohydrodynamic as well as electrohydrodynamic convection in tilted magnetic fields, patterns in stressed gel, and others.

**Acknowledgments**

This work was supported by the grant of the President of the Russian Federation no. MK-5801.2015.1 and also by the Ministry of education and science of the Russian Federation in the framework of the base part of the governmental ordering for scientific research works (project No 2479).

**Figures**

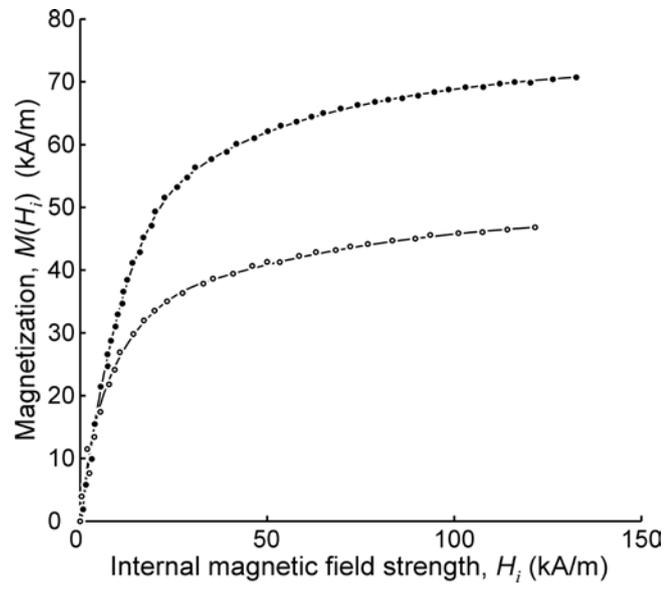

Fig. 1. Magnetization curves of the magnetic fluid used in experiment: MF-1 (filled symbols); MF-2 (open symbols). Lines are the approximation of measurement data.

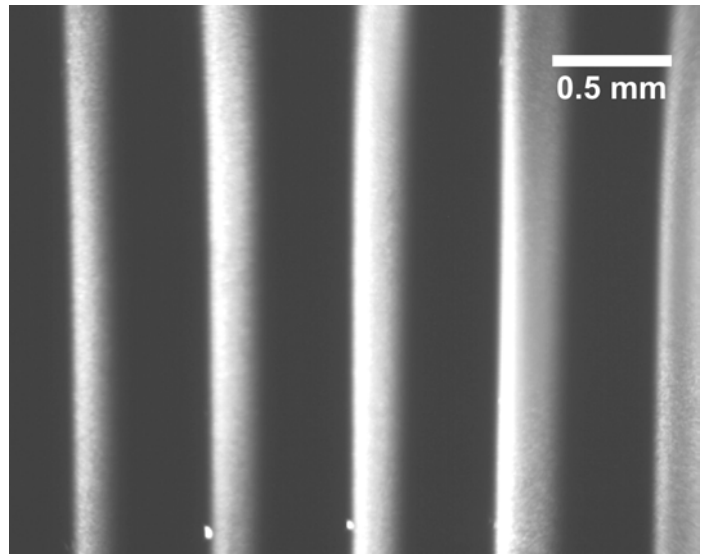

Fig. 2. Top view of the pattern of parallel liquid ridges arising from breakup of the magnetic fluid layer in a tilted magnetic field. Dark areas are the magnetic fluid, light areas are the glass substrate. The external magnetic field strength $H_e = 180$ kA/m, the layer thickness $h = 35$ μm, the tilt angle $\theta = 70°$. The scale bar is presented in the upper right corner.

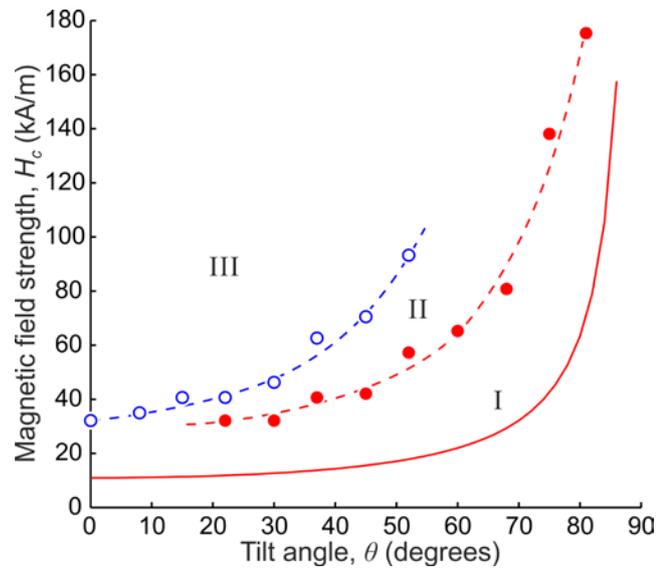

Fig. 3. Critical field strengths $H_{c1}$ (filled symbols) and $H_{c2}$ (open symbols) vs. the tilt angle, $\theta$. Layer thickness $h$ = 18 μm. Dots are experiments. Dashed lines are the approximation of experimental data; solid line is the calculations of $H_{c1}$. The numbers denote the surface configurations: I – flat surface, II – parallel ridges, III – conical peaks.

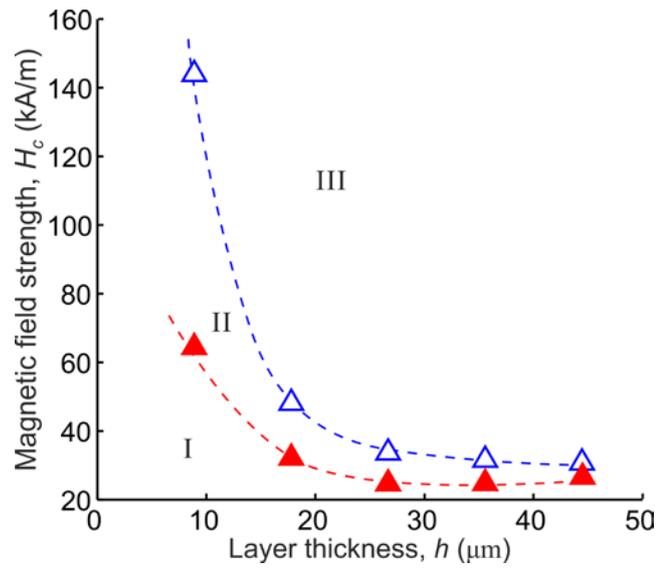

Fig. 4. The experimental dependencies of the first $H_{c1}$ (filled symbols) and the second $H_{c2}$ (open symbols) critical magnetic fields on the layer thickness at $\theta = 30°$. Lines are the approximation of experimental data. The numbers denote the surface configurations as in Fig. 3.

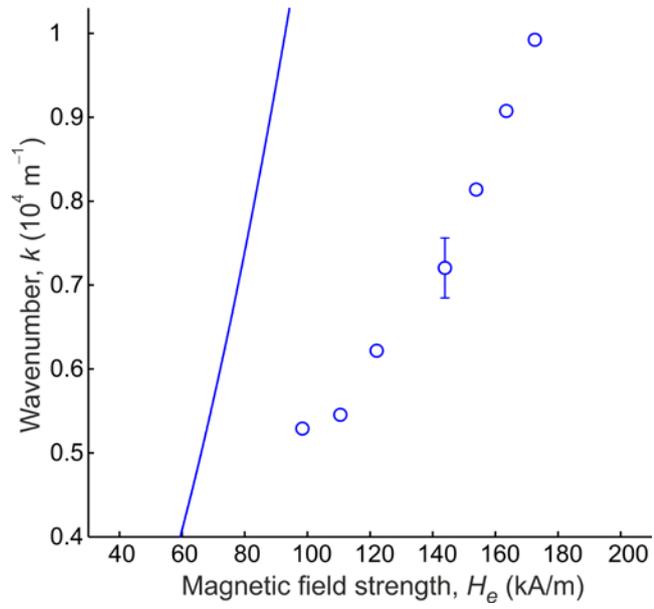

Fig. 5. The supercritical wave number, *k*, of the pattern of ridges vs. the external magnetic field strength, $H_e$, at $\theta = 70°$ and $h = 35$ μm. Dots give the experimental values; solid line shows the theoretical result.

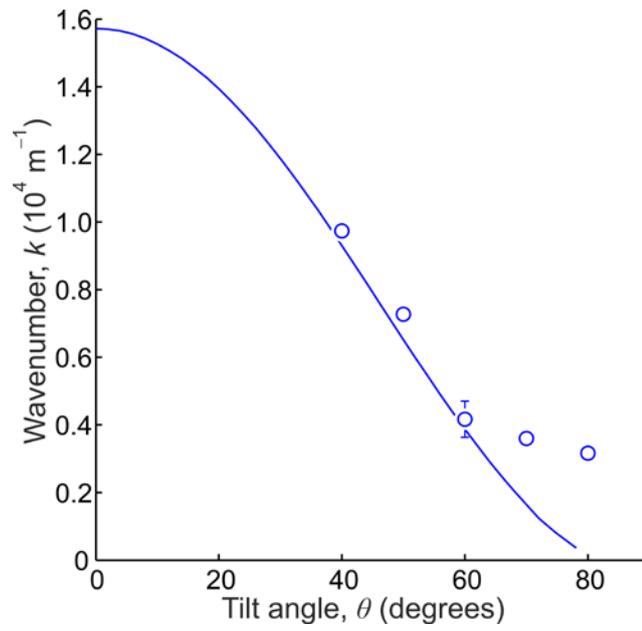

Fig. 6. The supercritical wave number vs. the tilt angle, $\theta$, at $H_e = 40$ kA/m and $h = 35$ μm. Dots are experiments; line is the calculations.

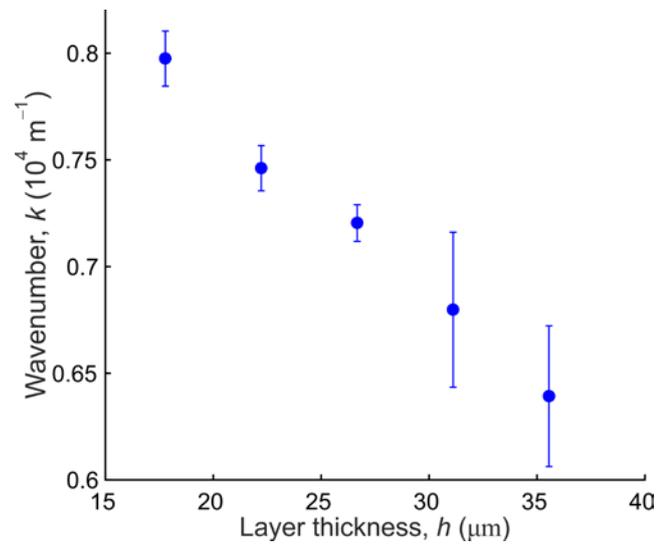

Fig. 7. The experimental dependence of the pattern wave number on the layer thickness at $H_e = 50$ kA/m and $\theta = 50°$.

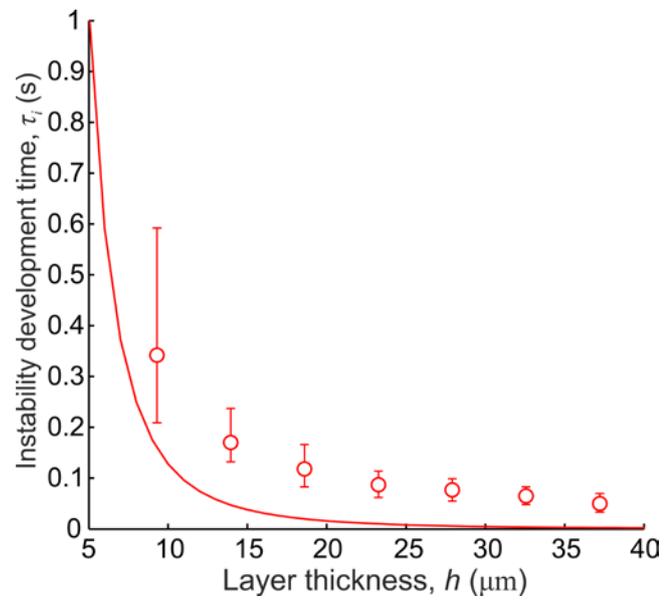

Fig. 8. The instability development time, $\tau_i$, vs. the layer thickness at $H_e = 60$ kA/m and $\theta = 0$. Dots are experiments; line is the calculations.

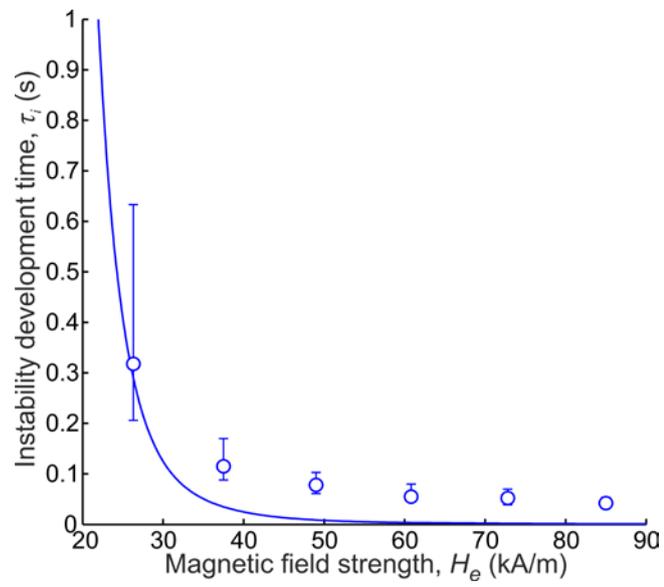

Fig. 9. The instability development time vs. the external magnetic field strength at $h = 33$ μm and $\theta = 0$. Dots are experiments; line is the calculations.